Modeling the effect of microstructure on elastic wave propagation in platelet-reinforced composites and ceramics

Hortense Le Ferrand,

School of Mechanical and Aerospace Engineering, School of Materials Science and Engineering, Nanyang Technological University, 50 Nanyang avenue, Singapore 639798

Corresponding email: hortense@ntu.edu.sg

*Dense ceramics are irreplaceable in applications requiring high mechanical stiffness, chemical and temperature resistance and low weight. To improve their toughness, ceramics can be reinforced with elongated inclusions. Recent manufacturing strategies have been developed to control the orientations of disc-like microparticles in polymeric and ceramic matrices and to build periodic microstructures. Given the infinite number of possible microstructures available, modeling tools are required to select the potentially best design. Periodic microstructures can be involved in elastic wave scattering to dissipate mechanical energy from vibrations. In this paper, a model is proposed to determine the frequency bandgaps associated to periodic architectures in composites and ceramics and the influence of microstructural parameters are investigated. The results are used to define guidelines for the future fabrication of hard bulk ceramic materials that combine traditional ceramic's properties with high vibration resistance.*

Ceramics are advantageous in many high technological applications such as turbine blades, pipelines or tiles of spacecrafts thanks to their chemical inertness, stability until high temperature and high strength and hardness. However, non-piezoelectric ceramic do not present the damping capacity that is required to dissipate the energy from high mechanical impacts and vibrations. Instead, microcracking will occur [1]. To prevent failure of structural parts submitted to vibrations and impacts, reinforced composite laminates are replacing fragile ceramics. However, polymer-based composites cannot sustain the temperatures and harsh environments experienced by turbine blades or space shuttles. Ceramic matrix composites (CMC) are ceramics reinforced with inclusions to increase their toughness and strength [2]. However, their vibration resistance should be further improved for long timer performance under high mechanical dynamic solicitations.

Reinforced polymer composite laminates with periodic arrangements have been found to display enhanced damping properties [3-5]. This property is also observed in biological composites with very low organic content such as in the dactyl club of the Mantis Shrimp [6]. In these composites, shock attenuation and energy dissipation is though to rise from concomitant phenomena: high hardness, local plastic deformation and frequency bandgaps in shear waves [7-9]. Periodic laminated structures in CMC can also be obtained [1-10] to augment the crack deflection path and the toughness. However, the onset of mechanical bandgaps in pure dense ceramics has not yet been explored. Furthermore, new methods have lead to the fabrication of dense multilayered composites and ceramics with high structural control at multiple levels [11,12]. The specimens fabricated consist in multilayered assemblies of predefined pitch and layer thickness, and with a 3-dimensional (3D) control over the orientation of disc-like inclusions within each layer. In contrast to long or short fiber reinforced composites, these platelet-reinforced materials present a larger variability of possible microstructures. So far, these architectured materials have been scarcely studied for mechanical wave dissipation. Notable papers relate the broadband absorption capacity in composites with a single orientation of platelets [13] and with two orientations [14]. Complex architectures such as those observed in biological composites [15-16] or fabricated synthetically [11,17,18] have not been explored so far experimentally nor theoretically.

In this paper, the effect of the microstructure of platelet-reinforced composites on the attenuation of elastic waves is studied theoretically using an analytical model. First, mechanical data available in the literature [19] are used to investigate the effect of the microstructure, layer thickness, platelet angle variation between layers and density on the frequency bandgaps appearing in the propagation of a normally-incident elastic wave. The findings are then applied theoretically to a fully dense ceramic system that has been experimentally fabricated in another work [12]. The results obtained in the platelet-reinforced composites and in the ceramic systems are then discussed in terms of combined static and dynamic mechanical performance, density and thermal resistance. The applicability of the modeling approach and its relevance for concrete experimental systems are also commented. The observations and the model developed in this study could present guidelines for the design of materials with new combinations of properties.

**Properties of epoxy composites reinforced with disc-like particles**

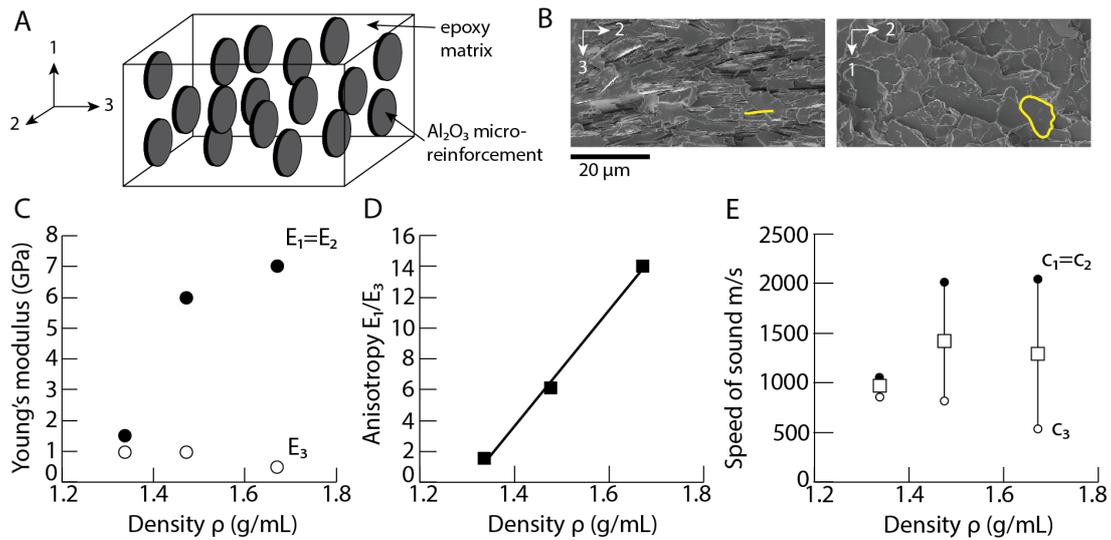

**Fig. 1. (A)** Schematics of a composite layer with aligned microplatelet reinforcement. **(B)** Electron micrographs showing the orthotropic direction of reinforcement in the plane (1-2) (images courtesy of Jascha Schmied). One microplatelet is highlighted in yellow in each micrograph. Characteristics of the material's aligned single element in function of the density $\rho$, namely **(C)** the Young's moduli along the axis 1 and 3, **(D)** the mechanical anisotropy defined as $E_1/E_3$ and **(E)** the materials' sound velocity along axis 1 and 3, namely $c_1$ and $c_3$ (data extracted from reference [20]).

Composites reinforced with biaxially aligned hard microplatelets have orthotropic characteristics [21]. In this study, literature data of alumina platelet reinforced epoxy composites are taken as a model system (fig.1). Reinforcement based on these microplatelets are been used in a large number of papers [18, 22-27] and the composites and ceramics fabricated have been extensively characterized by a variety of means including ultrasonic waves [19]. In the case considered here, the directions 1 and 2 are equivalent (see schematics fig. 1A). This biaxial alignment is obtained experimentally by vacuum pressing [28], shear [24,29] or rotating magnetic fields [11,13,30]. Electron micrographs of fractured cross-sections of these composites are reproduced in fig. 1B to highlight the orthotropy of the composite as well as the irregular shape of the platelets assimilated to discs.

Biaxial reinforcement increases the strength and toughness in monolayers [31] and can induce morphing in bilayers with perpendicular directions of reinforcement [19,32]. An increase in the volume fraction in reinforcing platelets is accompanied with an increase in the mechanical strength and stiffness of the composite [19,31] (fig. 1C). The anisotropy between the properties in the planes (1-2) and (1-3) increases alongside with platelet concentrations from 0 to 20% (fig. 1D). Despite the rising curve in fig. 1D, it can be expected that the anisotropy in Young's modulus will plateau at higher volume fractions, given the increase in concentration in the 3 directions of the materials with is accompanied with a lesser anisotropy in ultrasonic wave propagation velocity (fig. 1E). Higher volume fractions in reinforcing platelets can, however, be associated with other mechanical behaviors such as platelet-platelet friction and intragranular cracking [33,34] that can largely account for the augmentation in toughness observed in these materials [11,35,36].

**Modeling approach**

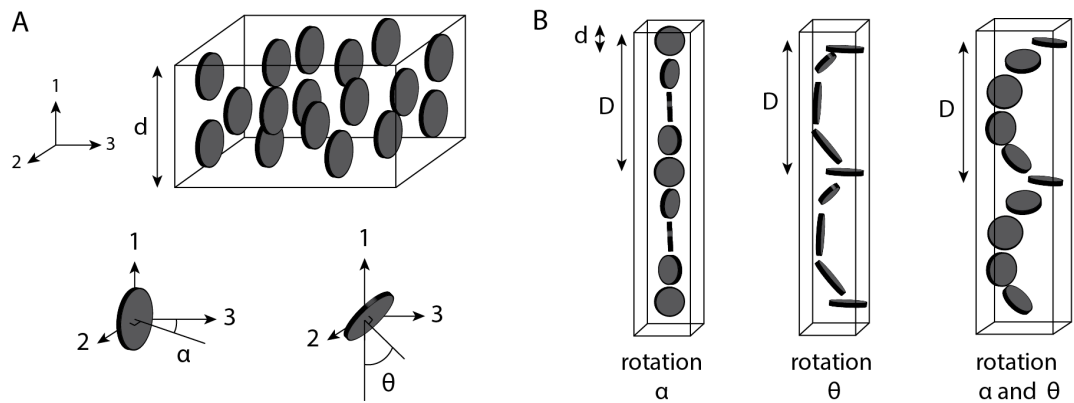

**Fig. 2. (A)** Schematics of a biaxially-reinforced orthotropic material's element and of the angles $\alpha$ and $\theta$ describing the platelet's orientation in 3D. **(B)** Description of the periodic microstructures modeled where $d$ is the thickness of one layer with one orientation of the platelets, $D$ the length of the period and $\alpha$ and $\theta$ the rotation angle of the platelets between consecutive layers.

The mechanical properties in orthotropic materials can be modeled with a stiffness tensor comprising 9 independent parameters. However, due to the biaxial reinforcement of each layer, two directions are equivalents, namely directions 1 and 2 (figure 2A). This therefore leads to following relations $E_1 = E_2$, $\nu_{13} = \nu_{23}$ and $G_{23} = G_{13}$, with $E_i$ the Young's modulus along the direction $i$, $\nu_{ij}$ the Poisson ratio and $G_{ij}$ the shear modulus in the plane $ij$ and reduces the number of independent parameters to 6. Contrary to long fiber composites where the fiber orientation is usually rotated around only one axis, the disc-like particles studied here can be rotated in the composite around 2 axis (2 axis are similar due to the disc geometry, see fig. 2A). Experimentally, this is realized using rotating magnetic fields [23]. The range of final microstructures available with variations in $\alpha$, $\theta$, $d$ and $D$ is thus infinite (fig. 2).

To model and evaluate whether periodic structures based on platelet reinforced composites and ceramics can attenuate an elastic wave by filtering out some of its content through the presence of bandgaps, three major periodic structures are considered (fig. 2B). First, the structure called rotation $\alpha$ is the Bouligand structure observed in biological composites [9, 37-39] and extensively reproduced in fibrous composites, and adapted to platelet inclusions. This structure has been hypothesized to filter shear waves [9]. Second, a structure called rotation $\theta$ will be modeled. Finally, a structure with mixed character will be considered, with rotations of angles $\alpha$ and $\theta$. Among the infinite pool of periodic structures imaginable, these three simple arrangements can provide the first guidelines to correlate the relationship between microstructure and high frequency impact filtering properties.

To this aim, an analytical model developed for fibers [7,9] is adapted to orthotropic layers of thickness $d$, assembled into infinitely periodic structures of period $D$. The incident wave impacting the material is an elastic wave entering the material with normal incidence. Rayleigh surface waves are not considered. The inverse of the stiffness tensor [C] of a layer organized as in fig. 2A is

$$[C]^{-1} = \begin{bmatrix} \frac{1}{E_1} & \frac{-v}{E_1} & \frac{-v}{E_3} & 0 & 0 & 0 \\ \frac{-v}{E_1} & \frac{1}{E_1} & \frac{-v}{E_3} & 0 & 0 & 0 \\ \frac{-v}{E_3} & \frac{-v}{E_1} & \frac{1}{E_3} & 0 & 0 & 0 \\ 0 & 0 & 0 & \frac{1}{G_{13}} & 0 & 0 \\ 0 & 0 & 0 & 0 & \frac{1}{G_{13}} & 0 \\ 0 & 0 & 0 & 0 & 0 & \frac{1}{G_{12}} \end{bmatrix}, \quad (1)$$

with $v$ the Poisson ratio estimated constant and equal to 0.25 as a first approximation, and therefore reducing the number of independent parameters down to 5, whereas $E_1$ and $E_3$ and $G_{13}$ and $G_{12}$ are the Young's and shear moduli determined experimentally by the ultrasonic method, respectively. In this layer, the tensor describing the propagation of a wave of frequency $\omega$ and wavenumber $k$ along the axis 1 is:

$$[P] = \begin{bmatrix} 0 & 0 & 0 & -i\frac{C_{44}}{\Delta} & -i\frac{C_{45}}{\Delta} & 0 \\ 0 & 0 & 0 & -i\frac{C_{45}}{\Delta} & -i\frac{C_{55}}{\Delta} & 0 \\ 0 & 0 & 0 & 0 & 0 & \frac{-i}{C_{33}} \\ i\rho\omega^2 & 0 & 0 & 0 & 0 & 0 \\ 0 & i\rho\omega^2 & 0 & 0 & 0 & 0 \\ 0 & 0 & i\rho\omega^2 & 0 & 0 & 0 \end{bmatrix} \quad (2)$$

with $\Delta = C_{44}C_{55} - C_{45}^2$ and $\rho$ the density of the material. The matrices $[P(\alpha,\theta)]$ for layers with rotated angles are obtained by applying the Bond transformations (see supplementary material for details) to the stiffness tensor. Then, a material constituted of multiple layers stacked into a complex structure is defined by the propagator matrix [Ps]:

$$[Ps]=i*\sum_{each\ layer} d \cdot [P(\alpha,\theta)], \quad (3)$$

The dispersion curves in the first Brillouin zone corresponding to a periodic microstructure are then obtained by solving the general eigenvalue problem arising from the Bloch theorem:

$$e^{[Ps]}V(0)=e^{-ikD}V(D)=e^{-ikD}V(0), \quad (4)$$

with V a state vector containing the displacement and the stresses propagating.

**Effect of the microstructure**

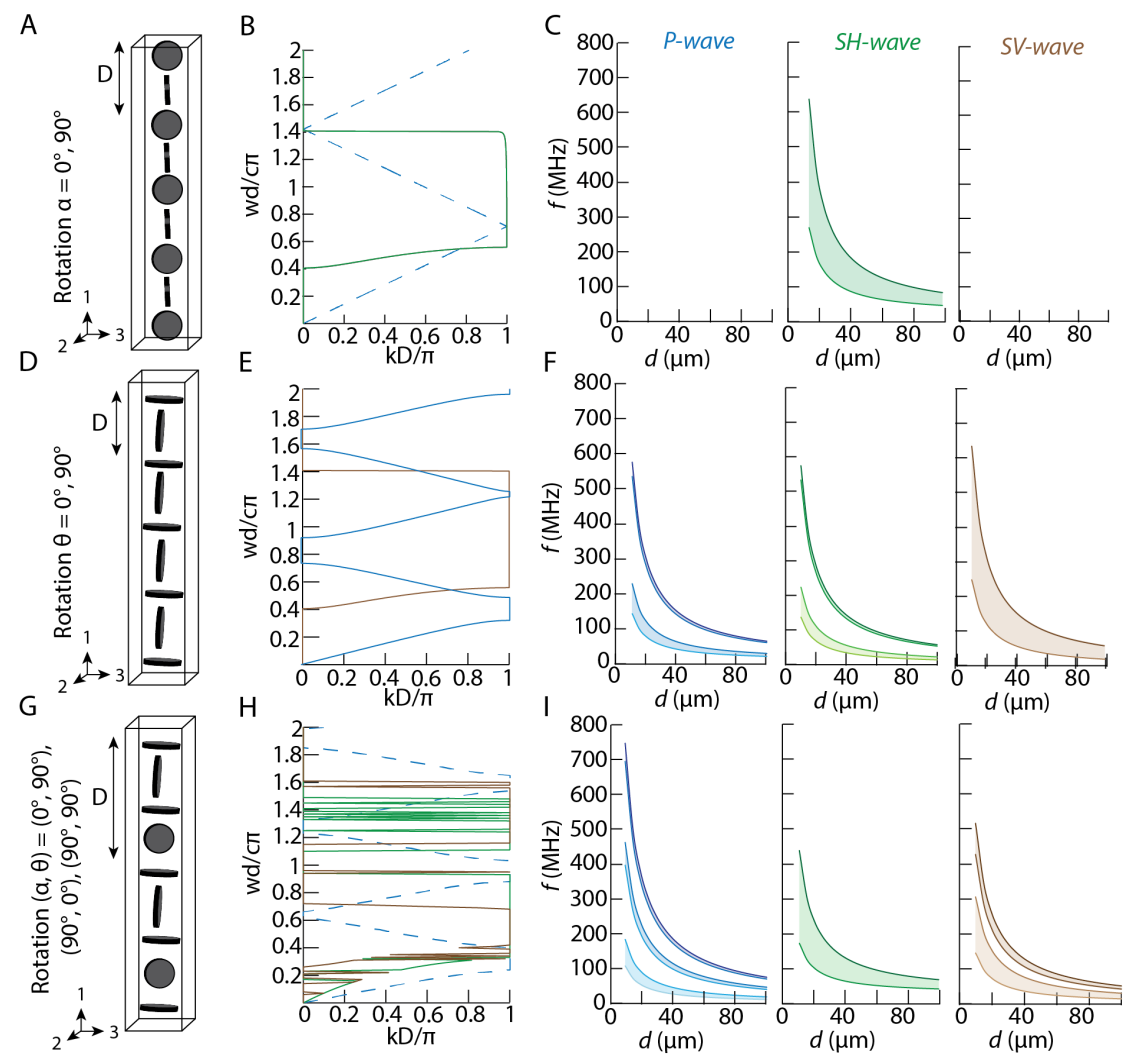

**Fig. 3. (A)** Schematics of a microstructure with rotation in angle $\alpha$ of 0 and 90°, **(B)** the corresponding dispersion curve in the first Brillouin zone and **(C)** the frequency bandgaps for normally incident compressive (P-, blue), shear (horizontally polarized, SH-, green, and vertically polarized, SV-, brown) elastic waves in function of the layer

thickness *d*. Same schematics and graphs for a microstructure with rotation of angle $\theta$ of 0 and 90° **(D-F)** and of angles $(\alpha, \theta)$ of values (0, 90), (90,0) and (90,90) **(G-I).**

The propagation of the incoming elastic wave is affected by the microstructure of the material it impacts. The case of epoxy composites reinforced with 10 vol% of alumina platelets and with the three periodic microstructures described in fig. 2B but with rotation angles of 90° between consecutive layers is presented in fig. 3. The frequencies of the bandgaps determined by the dispersion curves are calculated in function of the layer thickness *d* by:

$$f = \frac{n*c\pi}{d}, \tag{5}$$

where *c* is the average material's sound velocity and *n* the value in the dispersion curve at $kD/\pi = 1$.

The structures with rotations in $\alpha$ exhibit a bandgap only in the shear waves with horizontal polarization (SH-waves). Indeed, since the microstructure appears homogeneous along the vertical direction 1, the direction of impact, no dispersion is observed in the compressive P-waves and the shear waves with vertical polarization (SV-waves). The structures with rotations in $\theta$, however, are heterogeneous in all directions. The bandgaps the SV-waves in these structures are the largest and are similar to those for the SH-waves in the structures with rotation in $\alpha$. This is expected since the plane (1-2) is the rotation $\theta$ is equivalent to the plane (2-3) in the rotation $\alpha$. The mixed microstructure with rotations in $\alpha$ and $\theta$ also present bandgaps in the three propagation modes without any preferential wave attenuation.

Furthermore, the larger the layer thickness *d* the lower the frequencies of the bandgap. In highly demanding conditions such as those experienced during pyrotechnic shocks, vibrations up to the MHz can occur [40]. In the remainder of the paper, the layer thickness *d* will be taken as 200 µm. This thickness is also representative of experimental samples and can be fabricated by additive manufacturing techniques with magnetic orientation of platelets [11,19,26,41].

In hard materials, shear waves carry 3 to 4 times more energy than the compression waves. It is therefore primordial to design a microstructure with a large band gap in the shear modes, which is the case when the platelets are rotating in angle $\theta$. Since there is little difference between the microstructure with rotation in $\theta$ and in $(\alpha, \theta)$, further analysis will be carried on the structure with rotation in $\theta$.

## Effect of the angles

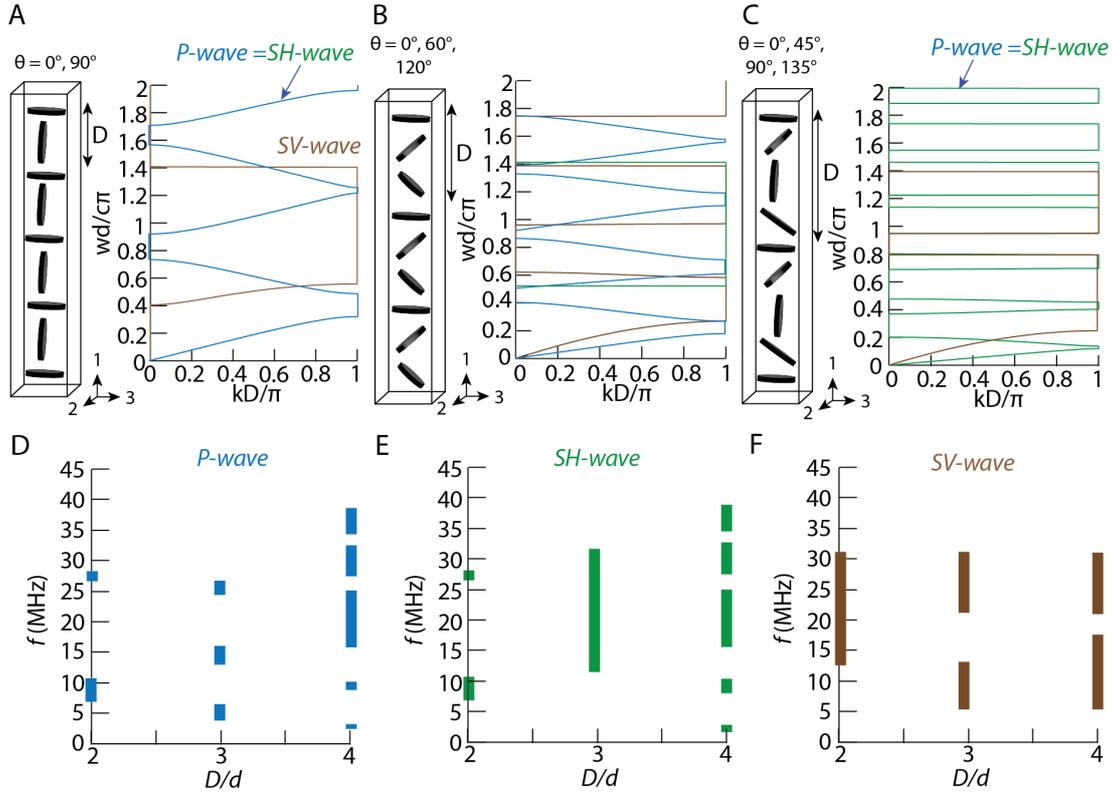

**Fig. 4.** Schematic and dispersion curves obtained for a microstructure with rotation in angle θ **(A)** 2 angles 0 and 90°, **(B)** 3 angles 0, 60 and 120°, and **(C)** 4 angles 0, 45, 90 and 135°. **(D), (E), (F)** represent the frequencies of the bandgaps for the P-wave, SH-wave and SV-wave, respectively, with a layer thickness $d$ of 200 μm.

Microstructures with angle rotations in θ exhibit the largest bandgaps in the shear mode of propagating elastic waves. These periodic microstructures can be realized in composite materials using magnetically-assisted slip-casting [11], a technique where the pitch, layer thickness, angle rotation step and density can be controlled. Applying the model to microstructures with rotations in θ with 2, 3 and 4 angle values, significant modifications in the bandgaps are observed (fig. 4). The frequencies of the bandgaps calculated for a layer thickness $d$ of 200 μm seem to increase with the number of angles. Furthermore, the bandgap in the compression P-waves are always narrower than for the shear waves with larger bandgaps in the vertically-polarized shear waves.

## Effect of the volume fraction in reinforcement

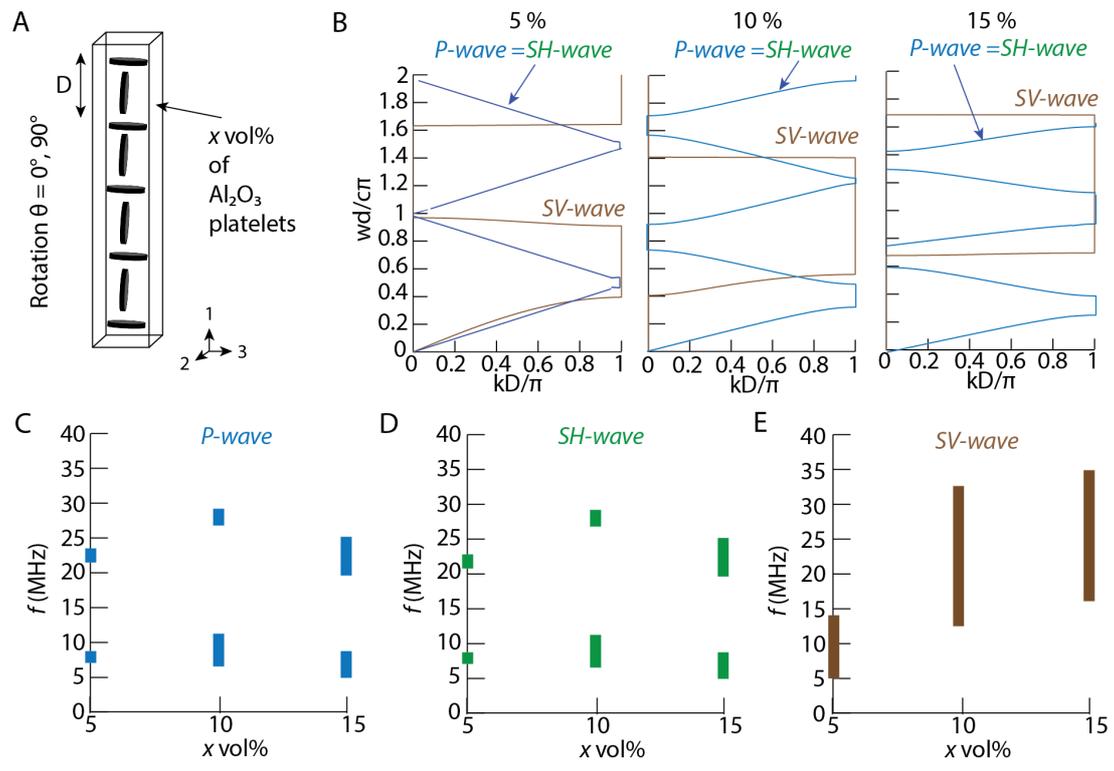

**Fig. 5. (A)** Schematics of the modeled architecture with *x* vol% of alumina platelets. **(B)** Dispersion curves obtained for increasing concentration in biaxially aligned alumina platelets. **(C-E)** Frequencies of the bandgaps in function of the volume fraction in platelets for the P-waves, SH-waves and SV-waves, respectively.

Polymeric composites can be reinforced with higher concentration of anisotropic platelets to increase their Young's modulus along the platelet's main plane direction. At the same time, the anisotropy in stiffness $E_1/E_3$ increases within the composite. Experimentally, the concentration in platelets is limited by the rise in viscosity during the fabrication process that hinders the alignment of the particles [31]. To study the effect of the increase in platelets' concentration on the frequency bandgap, the bandgaps are determined for a microstructure in rotation $\theta$ with two angles. An increase in particle concentration leads to increase in the frequency bandgap and frequency range of SV-waves but decreases the frequencies of filtered P and SH-waves (fig. 5). However, at low concentrations in particles, the effect is opposite for the P and SH-wave and is likely to be related to the decrease in the materials' sound velocity (fig. C, D). From these results, it can nevertheless be expected that polymeric composites with higher platelets concentration could filter shear waves over larger frequencies. However, this is without considering the decrease in anisotropy likely to occur past a certain concentration as all directions are reinforced.

## Application to microstructured ceramics

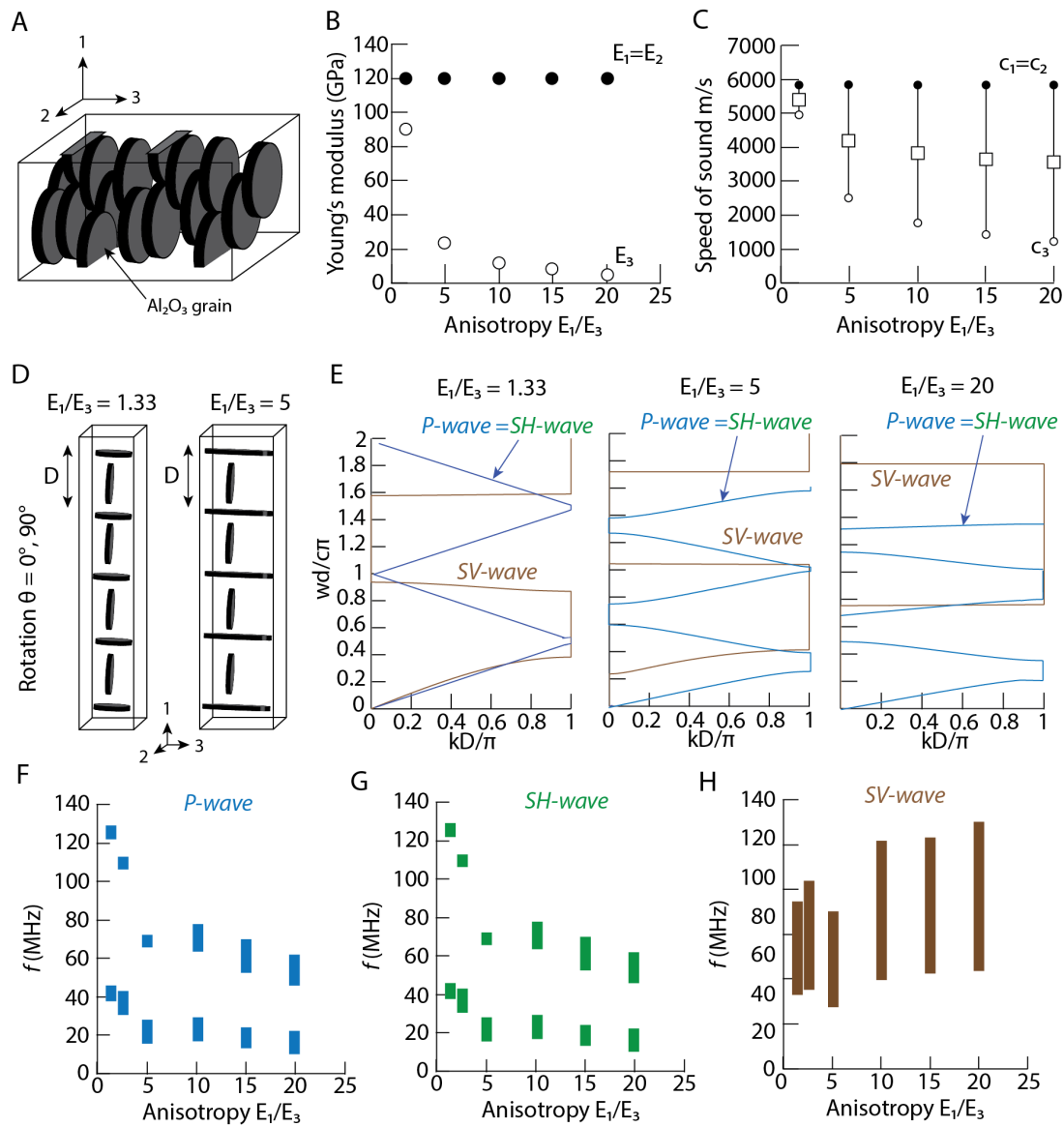

**Fig. 6. (A)** Schematics of an orthotropic layer of alumina ceramic with aligned grains. **(B)** and **(C)** are the Young's modulus and the materials' speed of sound from an experimental sample [12] (anisotropy 1.33) and estimated for increased anisotropy. **(D)** Schematics of the microstructure modeled for the bandgap determination and **(E)** the dispersion curves obtained for three values of anisotropy. **(F), (G)** and **(H)** are the frequency bandgaps in the first Brillouin zone for the three propagating modes and for a layer thickness $d$ of 200 µm.

Dense ceramics with deliberate microstructure have been obtained by using a templated grain growth process where microplatelets oriented in space with magnetic fields in suspension of nanoparticles dictate the orientation of the ceramic anisotropic grains after sintering [12,42]. Periodic microstructures were fabricated either by layering slurries of different compositions [10] or by tuning the orientation of the grains using time-dependent magnetic fields [12]. Samples produced using this latter process exhibit local control over the mechanical properties and a large flexibility in the orientations and thicknesses achievable. Based on the mechanical properties measured via the ultrasound method, the mechanical data are used as input in the model (fig. 6A-C). Since these are dense ceramics without a secondary phase, the parameter that can be tuned in addition to the platelets angles ($\alpha, \theta$) and layer thickness *d*, is the anisotropy in mechanical properties. Modeling the effect of an increase in anisotropy could be used to define new guidelines for the fabrication of periodic ceramics with specific bandgaps. The range in anisotropy chosen varies from 1.33 to 20, where 1.33 corresponds to the real experimental sample [12], whereas 20 is a realistic value for templated aluminas with elongated grains obtained using doping ions [43]. The microstructures with rotations of angle $\theta$ from 0 to 90° and increasing anisotropy, the dispersion curves have different profiles (fig. 6D-E). Generally, the bandgaps in the P and SH waves decrease in frequency but increase in width as the anisotropy increases. The bandgap in SV wave, on the contrary, increases in the frequency range. To fabricate a ceramic with efficient mechanical vibration absorption, it seems therefore preferable to have a ceramic with large anisotropy in modulus between the two directions 1 and 3.

## Strong ceramics with frequency bandgaps

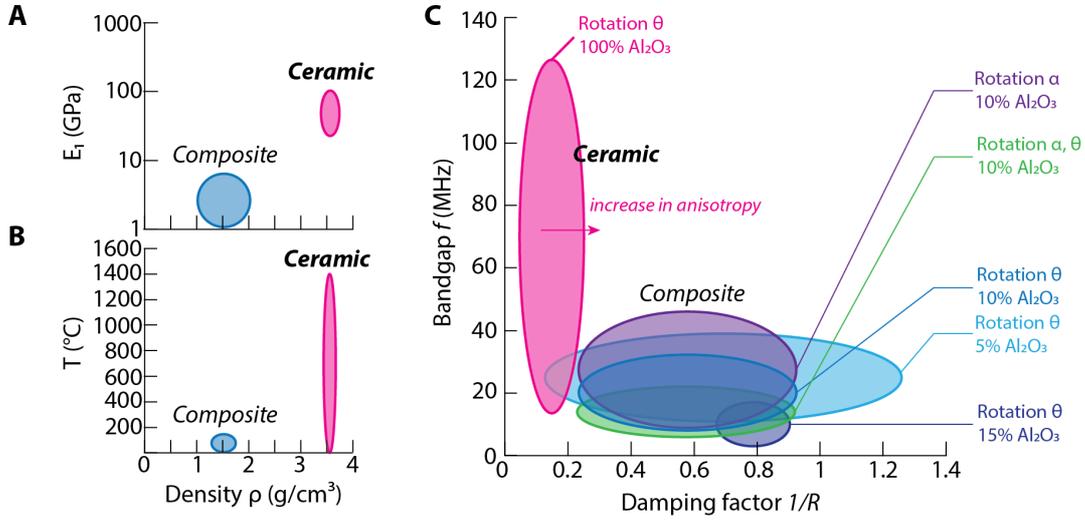

**Fig. 7.** Ashby-like plots of alumina microplatelet-reinforced epoxy composites and ceramics representing (A) the elastic modulus $E_1$ and (B) the temperature of operation $T$ in function of the density $\rho$, and (C) the bandgap in function of the damping factor $1/R$. In (C), the bandgaps were calculated for rotations of 0 and 90° of angles $\alpha$, $\theta$ or $(\alpha, \theta)$.

Ceramics exhibit high elastic moduli and operating temperatures in comparison with polymer-based composites. Tuning the microstructure within bulk ceramics is predicted to be a means to artificially attenuate high velocity elastic waves over a large frequency range (fig. 7).

Reinforcing thermoset matrices with ceramic particles increases to a limited extend their elastic modulus, but does not increases the operating temperature of the materials. Typically, above 100°C, the glass transition temperature reduces dramatically the strength of these materials that then decompose above 450 °C. In contrast, ceramics exhibits elastic modulus at least one order of magnitude higher than the polymer composites and can be operational up to 1300 °C (fig. 7AB).

However, composite materials benefit from the viscoelastic properties of the polymer matrix to dissipate mechanical energy. This dissipation can be represented by $\frac{1}{R}$ where R is the radiation coefficient [44]:

$$R = \frac{c}{\rho} = \sqrt{\frac{E}{\rho^3}}, \tag{6}$$

with *c* the material's speed of sound, *E* its Young modulus and *ρ* its density. Since the materials considered here are highly heterogeneous due to the microstructuring, average values in *E* and *c* are taken to calculate *R*.

The model described in this study thus predicts large frequency bandgaps over the MHz range for periodically microstructured ceramics with rotation in $\theta$ and a layer thickness of 200 µm. Larger thicknesses would decrease the frequency range to achieve kHz. Increasing the anisotropy increases the apparent dissipation effect by reducing the Young's modulus in one direction. Another experimental means to increase the anisotropy along with the damping in dense ceramics could be to incorporate solid lubricant such as graphene flakes in the formulations, and to orient them along with the grains. Studies have shown that the addition of graphene flakes or reduced graphene oxide at wt% below 15 increases the anisotropy in young's modulus of bulk silicon-nitride ceramics [45]. Also, graphene is actively involved in multiple toughening phenomena [46,47]. Finally, employing piezoelectric compositions [48,49], could further increase the dissipation of vibrational energy by heat conversion.

**Discussion and conclusion:**

The analytical model traditionally used for transversally isotropic periodic laminates was adapted to orthotropic microperiodic structures with biaxial reinforcements. Experimental mechanical characteristics were used as input to determine the effect of the microstructural parameters on the frequency bandgaps in platelet-reinforced epoxy composites, namely the platelet rotation angles, the layer thickness, the number of layers in each period and the concentration in reinforcement. The findings show that the microstructure with rotation of angles $\alpha = 0$ and 90° have the largest bandgaps in shear waves SV. The composition and the platelet volume fraction directly impact the frequency range. A dimensional analysis using the relation $f = c/\lambda$ indeed shows that at these frequencies, the waves are likely to interact with the interparticles' distances ranging from 10 to 100 µm (see figure in supplementary material). However, to be allowed to consider each layer of biaxial reinforcement as a homogeneous orthotropic material, for which only the modeling carried out here is valid, the ratio between $\lambda$ and the particulate characteristic length should be sufficiently large, which is the case for frequencies larger than 150 MHz.

Applying the model to dense microstructured periodic ceramics, design guidelines can be extracted to create materials that exhibit high stiffness, high temperature resistance and that can filter selected ranges of frequencies. These materials could be expected to exhibit superior resistance the intense vibration like those generated by pyrotechnic shocks. The conclusions from the model is that microstructures with alternating layers with particles orientations in 0 and 90° with layer thickness ~200 μm and with large anisotropies in grain sizes are likely to behave as desired.

The model presented here is anticipated to be used as a tool to select candidate microstructures from the pool of infinite possible structures available for these materials. This study sets a first homogenization scheme to predict the response of periodic materials with biaxially reinforced orthotropic layers. To validate the results of this modeling, further experimental characterization should be carried out to estimate the non-linear effects occurring within the materials, that are not captured in this homogenization, such as friction platelet-matrix and platelet-platelet, local strain and stresses, local heating… Selected designs could be fabricated and tested to verify this first approximation. Furthermore, combining multiple designs could then be envisaged to generate broadband wave attenuation, for example by juxtaposition of a periodic polymeric composite as a baking layer to a periodic ceramic. Also, similar principles could be applied to attenuate Rayleigh waves traveling at the surface of the impacted material by building a periodic material along the directions 2 and 3.

**Methods:**

**Numerical Solutions**

The eigenvalue problem (4) was solved using MATLAB (MathWorks, Inc., USA), taking in input the experimental values as provided in the literature. The Bond transformation tensors are detailed in the supplementary materials.

**Data availability**

All data generated or analyzed in this study are included in this published article.

**Acknowledgment**

Florian Bouville is acknowledged for discussions and inputs and ETH Zürich, Switzerland for the MATLAB license. Nanyang Technological University start-up grant M4082382.050 is acknowledged.

**Competing interests**

The author declares no competing interests.

**Author contribution**

Modeling the effect of microstructure on elastic wave propagation in platelet-reinforced composites and ceramics

Hortense Le Ferrand,

School of Mechanical and Aerospace Engineering, School of Materials Science and Engineering, Nanyang Technological University, 50 Nanyang avenue, Singapore 639798

Corresponding email: hortense@ntu.edu.sg

**Supplementary Material**

**Bond transformation tensors:**

The stiffness tensors of rotated layers are determined using the Bond transformation:

The matrix describing a clockwise rotation of angle $\alpha$ is:

$$[\text{rotz}(\alpha)] = \begin{bmatrix} \cos(\alpha) & \sin(\alpha) & 0 \\ -\sin(\alpha) & \cos(\alpha) & 0 \\ 0 & 0 & 1 \end{bmatrix}.$$

and the Bond transformation matrices are:

$$[\text{Bondstress}(\alpha)] = \begin{bmatrix} \cos^2(\alpha) & \sin^2(\alpha) & 0 & 0 & 0 & \sin(2\alpha) \\ \sin^2(\alpha) & \cos^2(\alpha) & 0 & 0 & 0 & -\sin(2\alpha) \\ 0 & 0 & 1 & 0 & 0 & 0 \\ 0 & 0 & 0 & \cos(\alpha) & -\sin(\alpha) & 0 \\ 0 & 0 & 0 & \sin(\alpha) & \cos(\alpha) & 0 \\ -\frac{\sin(2\alpha)}{2} & \frac{\sin(2\alpha)}{2} & 0 & 0 & 0 & \cos(2\alpha) \end{bmatrix}$$

and

$$[\text{Bondstrain}(\alpha)] = \begin{bmatrix} \cos^2(\alpha) & \sin^2(\alpha) & 0 & 0 & 0 & \frac{\sin(2\alpha)}{2} \\ \sin^2(\alpha) & \cos^2(\alpha) & 0 & 0 & 0 & \frac{-\sin(2\alpha)}{2} \\ 0 & 0 & 1 & 0 & 0 & 0 \\ 0 & 0 & 0 & \cos(\alpha) & -\sin(\alpha) & 0 \\ 0 & 0 & 0 & \sin(\alpha) & \cos(\alpha) & 0 \\ -\sin(2\alpha) & \sin(2\alpha) & 0 & 0 & 0 & \cos(2\alpha) \end{bmatrix}$$

The stiffness tensor [L] for a rotation of angle $\alpha$ is therefore calculated following:

$[L] = [\text{Bondstress}(\alpha)][C][\text{Bondstrain}(\alpha)]^{-1}$

$$[L] = \begin{bmatrix} L_{11} & L_{12} & L_{13} & 0 & 0 & L_{16} \\ L_{12} & L_{22} & L_{21} & 0 & 0 & L_{26} \\ L_{13} & L_{23} & C_{11} & 0 & 0 & L_{36} \\ 0 & 0 & 0 & C_{66}\cos^2(\alpha) + C_{44}\sin^2(\alpha) & (C_{66} - C_{44})\cos(\alpha)\sin(\alpha) & 0 \\ 0 & 0 & 0 & (C_{66} - C_{44})\cos(\alpha)\sin(\alpha) & C_{44}\cos^2(\alpha) + C_{66}\sin^2(\alpha) & 0 \\ L_{16} & L_{26} & L_{36} & 0 & 0 & L_{66} \end{bmatrix}.$$

The matrix describing a clockwise rotation of angle $\theta$ is:

$$[\text{rotx}(\theta)] = \begin{bmatrix} 1 & 0 & 0 \\ 0 & \cos(\theta) & \sin(\theta) \\ 0 & -\sin(\theta) & \cos(\theta) \end{bmatrix}.$$

and the Bond transformation matrices are:

$$[\text{Bondstress}(\theta)] = \begin{bmatrix} 1 & 0 & 0 & 0 & 0 & 0 \\ 0 & \cos^2(\theta) & \sin^2(\theta) & \sin(2\theta) & 0 & 0 \\ 0 & \sin^2(\theta) & \cos^2(\theta) & -\sin(2\theta) & 0 & 0 \\ 0 & \frac{-\sin(2\theta)}{2} & \frac{\sin(2\theta)}{2} & \cos(2\theta) & 0 & 0 \\ 0 & 0 & 0 & 0 & \cos(\theta) & -\sin(\theta) \\ 0 & 0 & 0 & 0 & \sin(\theta) & \cos(\theta) \end{bmatrix}$$

and

$$[\text{Bondstrain}(\theta)] = \begin{bmatrix} 1 & 0 & 0 & 0 & 0 & 0 \\ 0 & \cos^2(\theta) & \sin^2(\theta) & \frac{\sin(2\theta)}{2} & 0 & 0 \\ 0 & \sin^2(\theta) & \cos^2(\theta) & \frac{-\sin(2\theta)}{2} & 0 & 0 \\ 0 & -\sin(2\theta) & \sin(2\theta) & \cos(2\theta_n) & 0 & 0 \\ 0 & 0 & 0 & 0 & \cos(\theta) & -\sin(\theta) \\ 0 & 0 & 0 & 0 & \sin(\theta) & \cos(\theta) \end{bmatrix}$$

The stiffness tensor [L'] for a rotation of angle $\theta$ is calculated following:

[L'] =[Bondstress($\theta$)][C][Bondstrain($\theta$)] $^{-1}$

$$[L'] = \begin{bmatrix} L'_{11} & L'_{12} & L'_{13} & 0 & 0 & L'_{16} \\ L'_{12} & L'_{22} & L'_{23} & 0 & 0 & L'_{26} \\ L'_{13} & L'_{23} & L'_{33} & 0 & 0 & L'_{36} \\ 0 & 0 & 0 & L'_{44} & 0 & 0 \\ 0 & 0 & 0 & 0 & L'_{55} & 0 \\ L'_{16} & L'_{26} & L'_{36} & 0 & 0 & L'_{66} \end{bmatrix}, \text{ where}$$

$L'_{33} = C_{11} \sin^4(\theta) + C_{33} \cos^4(\theta) + 2\cos^2(\theta)\sin^2(\theta)(C_{13} + 2C_{44})$

$L'_{44} = C_{44}(2\cos^2(\theta) - 1)^2 + (C_{11} - 2C_{13} + C_{33})\cos^2(\theta)\sin^2(\theta)$

$L'_{55} = C_{44}\cos^2(\theta) + C_{66}\sin^2(\theta)$

**Dimensional analysis:**

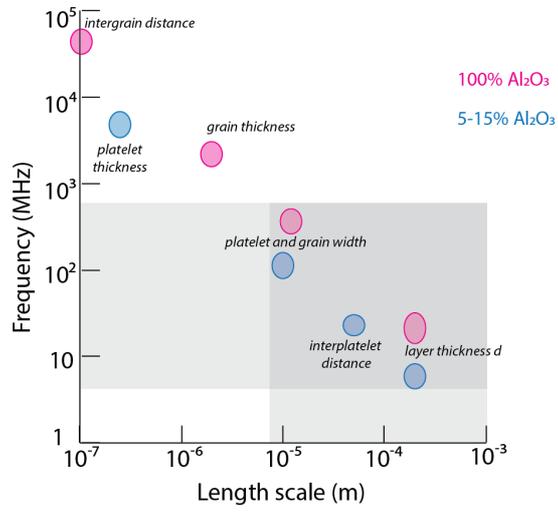